\documentclass[amsmath,superscriptaddress,prb,twocolumn] {revtex4}
\usepackage{bm}
\usepackage{enumerate}
\usepackage{graphicx}
\usepackage[dvips]{epsfig}
\usepackage{epsf}
\usepackage{xcolor}

\makeatletter
\newcommand{\Rmnum}[1]{\expandafter\@slowromancap\romannumeral #1@}
\makeatletter

\begin{document}
\title{Magnon thermal Hall effect in collinear antiferromagnets}
\author{Vladimir A. Zyuzin}
\affiliation{L.D. Landau Institute for Theoretical Physics, 142432, Chernogolovka, Russia}
\begin{abstract}
In this paper we theoretically discuss thermal Hall effect of magnons in insulating N\'{e}el ordered antiferromagnets at zero external magnetic field. 
We show that for compensated N\'{e}el order the non-zero thermal Hall effect will occur in the absence of any symmetry between the two magnetic sublattices, thus making the system ferrimagnetic. We then show that collinear Dzyaloshinskii's weak ferromagnets, in which there is a symmetry connecting the magnetic sublattices, also show magnon thermal Hall effect. The thermal Hall effect of magnons will be non-zero by a virtue of the spin-momentum splitting of the magnon spectrum due to the Dzyaloshinskii-Moriya interaction as well as second-nearest exchange interaction different in the two magnetic sublattices, both corresponding to the broken symmetries that lead to the Dzyaloshinskii's invariant. We construct a theoretical model in which an external electric field may change the symmetry of the antiferromagnetic system thus altering the thermal Hall effect of magnons. 
\end{abstract}
\maketitle

\section{Introduction}
There are two types of antiferromagnets in the classification scheme made by Turov \cite{Turov1965,Turov1990} of antiferromagnets with equal (connected by a symmetry operation) magnetic sublattices. Namely, these are genuine antiferromagnets with zero net magnetization and Dzyaloshinskii's weak ferromagnets \cite{BorovikRomanov1957,Dzyaloshinskii1958} with small non-zero magnetic moment. In Dzyaloshinskii's weak ferromagnets crystal symmetry and special directions of the N\'{e}el vector allow for the magnetic moment. In addition, there are antiferromagnetic ferrimagnets in which case magnetic sublattices with equal in magnitude and antialigned spins aren't connected by any symmetry operation.

In addition, more insight comes from the point of view of symmetries of the magnetic unit cell of an antiferromagnet.
For example, the local non-magnetic environment of the N\'{e}el order in metallic antiferromagnets may give rise to spin splitting of conducting fermions that interact with the N\'{e}el order \cite{Noda2016,Okugawa2018,HayamiYanagiKusunose2019,Naka2019,AHE_AFM,Ahn2019,Rashba2020,HayamiYanagiKusunose2020,Zyuzin2025}.
The spin-splitting may be antiferromagnetic or ferromagnetic depending on what symmetries the non-magnetic environment breaks. If the non-magnetic environment is such that the magnetic sublattices are connected by a combination of time-reversal and translation or inversion, then the conducting fermions are spin degenerate with zero net magnetic moment.
If the non-magnetic environment is such that only a combination of certain rotations or mirror operations and time-reversal connects the magnetic sublattices, then the spin-splitting of conducting fermions will be of the antiferromagnetic type corresponding to the rotation or mirror symmetry. Such spin-splittings might be of the $d-$,$g-$,$i-$ types \cite{Noda2016,Okugawa2018,HayamiYanagiKusunose2019,Naka2019,AHE_AFM,Ahn2019,Rashba2020,HayamiYanagiKusunose2020} or mirror-symmetric \cite{Zyuzin2025}. The d-wave spin splitting has been experimentally observed in Ref. \cite{KharchenkoBibikEremenko1985,EremenkoKharchenko1987} through the d-wave Hall effect \cite{VorobevZyuzin2024}. It was shown that the spin-splitting together with spin-orbit coupling may result in the anomalous Hall effect \cite{TurovShavrov,Turov1990,AHE_AFM} in antiferromagnets. Such antiferromagnets are Dzyaloshinskii's weak ferromagnets in which the conducting fermions are responsible for magnetization while the N\'{e}el order is intact.

In the case when the local non-magnetic environment breaks all symmetries between the magnetic sublattices, the spin-splitting of conducting fermions will acquire a ferromagnetic contribution which would result in finite spin-polarization of conducting fermions \cite{Naka2020,Zyuzin2024,Zyuzin2025}. Such antiferromagnets are called ferrimagnets. In a ferrimagnet studied in \cite{Zyuzin2025} the spin-splitting of conducting fermions was a sum of antiferromagnetic mirror-symmetric and ferromagnetic terms. It was shown that ferrimagnets have an anomalous Hall effect \cite{Zyuzin2025}.

It is then natural to understand implications of the local non-magnetic environment of the N\'{e}el order in insulating antiferromagnets on the thermal Hall effect (THE) of magnons \cite{MatsumotoShindouMurakamiPRB2014,LaurellFiete,Mook2019,ZyuzinKovalevPRL2016,Pires,Mook} in the absence of external magnetic field.
In \cite{ZyuzinKovalevPRL2016} the THE of magnons was shown to be absent in the collinear antiferromagnet on a honeycomb lattice in which case the two sublattices are connected by a combination of time-reversal and $\frac{\pi}{3}$ rotation or by a combination of time-reversal and inversion. In \cite{Pires} antiferromagnetic magnons have been studied in collinear antiferromagnet on a checkerboard lattice in which case the two sublattices are connected by a combination of time-reversal and $\frac{\pi}{2}$ rotation. The THE of magnons is absent in this system as well because of the symmetry between the magnetic sublattices.
In \cite{Mook,CommentAlter} a three-dimensional antiferromagnet on a rutile lattice was studied and it was found that non-zero THE of magnons is due to first-nearest neighbor Dzyaloshinskii-Moriya interaction. The system studied in \cite{Mook,CommentAlter} belongs to the Dzyaloshinskii's weak ferromagnet type \cite{Dzyaloshinskii1958,Turov1965}, hence the obtained non-zero magnon THE.

In this paper we wish to systematically understand the mechanism of magnon THE in insulating collinear antiferromagnets. By collinearity we mean absence of any canting of the N\'{e}el order.
We study theoretical two-dimensional model of an antiferromagnet proposed in \cite{Zyuzin2025}. This model allows to tune between different antiferromagnetic phases (genuine antiferromagnet, ferrimagnet and weak ferromagnet) in a simple way.
We deduce Dzyaloshinskii's invariants of existence of a finite magnetic moment in collinear antiferromagnets for ferrimagnet and weak ferromagnet models and show that the derived by us magnon THE in the models is consistent with the invariant. 
We first show that finite magnon THE can be found in insulating collinear antiferromagnets in which there is no symmetry between the two magnetic sublattices, i.e. in ferrimagnets. 
Absence of symmetry is due to the non-symmetric environment of the non-magnetic atoms. In addition to this, the first-nearest neighbor Dzyaloshinskii-Moriya interaction (DMI) consistent with the asymmetry of the magnetic sublattices as well as the second-nearest Heisenberg exchange interaction (HEI) different in the two magnetic sublattices are required for the non-zero magnon THE. Both ingredients are consistent with the Dzyaloshinskii's invariant.
We then show that magnon THE indeed appears in a weak ferromagnetic phase of the model proposed in \cite{Zyuzin2025}. Weak ferromagnet is a collinear antiferromagnet in which magnetic sublattices are connected to each other by some symmetry operation and which allow for the finite magnetic moment. The magnon THE in the weak ferromagnetic phase is also governed by DMI and second-nearest HEI consistent with the lattice symmetry and Dzyaloshinskii's invariant.

Furthermore, we suggest that the external electric field can vary the symmetry of the lattice when the unit cell of the insulating antiferromagnet can have a finite electric dipole moment. The system can then be driven into a low symmetry phase by the electric field. In that way, by varying the direction of the field, the magnon THE can be altered. This can serve as a powerful experimental tool to probe the structure of antiferromagnetic magnons.

\begin{figure}[t] 
\includegraphics[width=0.4 \columnwidth ]{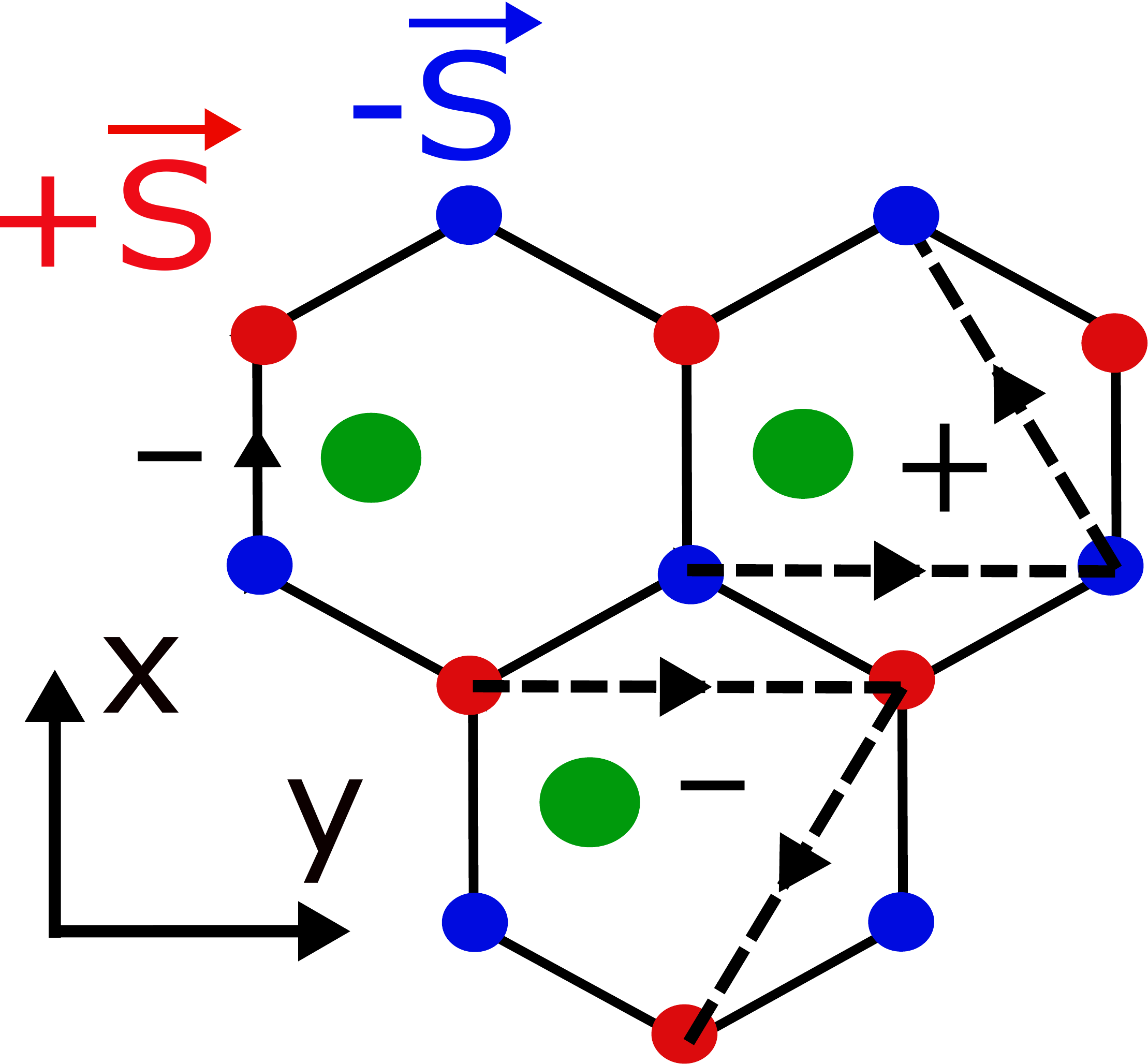} ~~~~
\includegraphics[width=0.4 \columnwidth ]{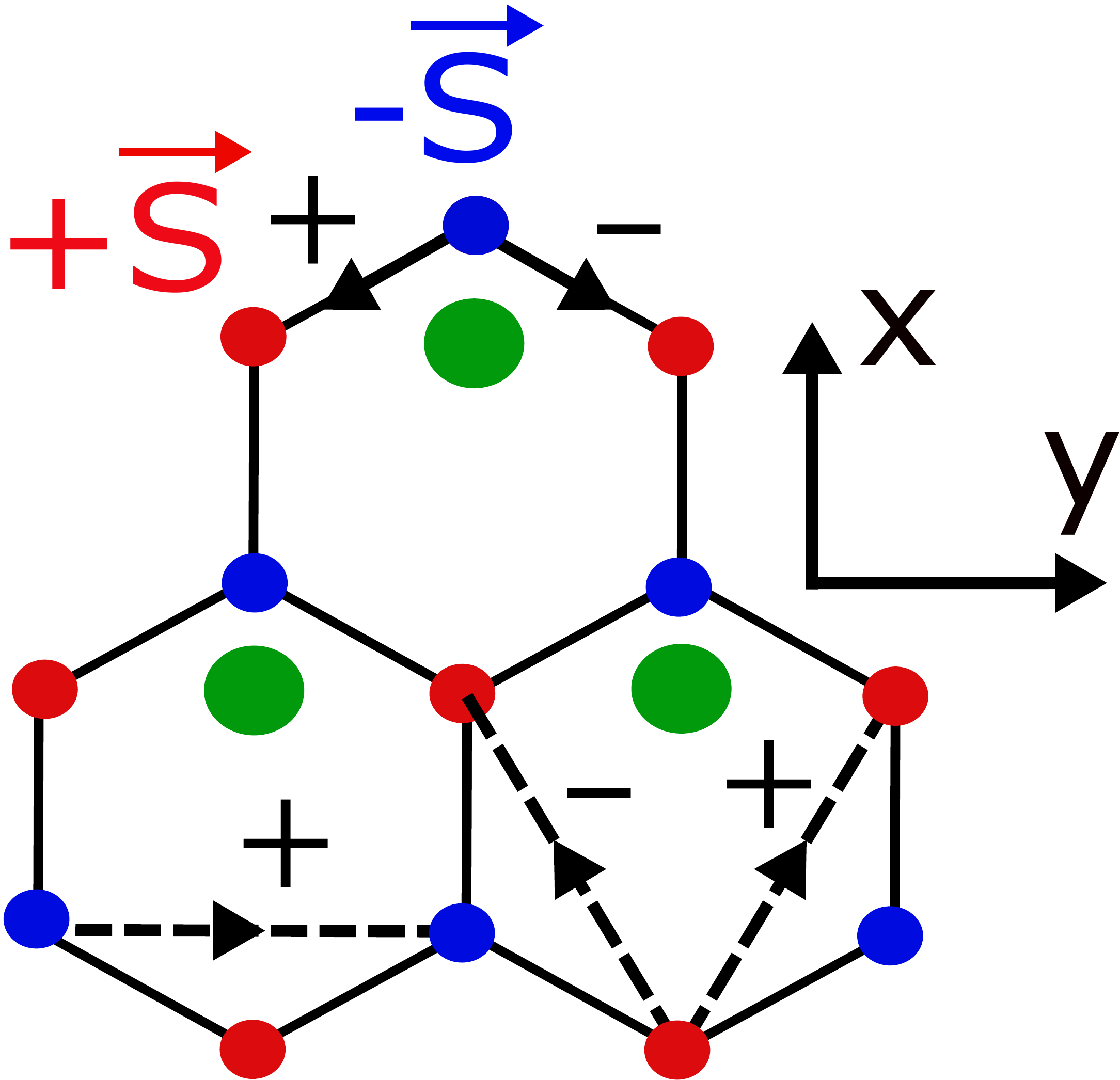} ~~

\protect\caption{The models of two antiferromagnetic systems are depicted. The red and blue sites represent the spin-up and spin-down magnetic sublattices of the N\'{e}el order. The green atom is non-magnetic and it is placed strictly in the plane with the red and blue sites.
The systems include a nearest-neighbor Heisenberg exchange interaction (HEI, $J>0$) and Dzyaloshinskii-Moriya interaction (DMI). Further-neighbor HEI and DMI are indicated by dashed lines. In a simplifying assumption, we exclude any interactions across the green atom.
The arrows on the bonds denote the sign of the DMI. The left model possesses a symmetry connecting the two magnetic sublattices, whereas this symmetry is absent in the right model.}
\label{fig:fig1}  
\end{figure}

\section{Model}
We study two insulating antiferromagnetic systems with the N\'{e}el order on the honeycomb-like lattices shown in Fig. (\ref{fig:fig1}). These models were studied in \cite{Zyuzin2025} for the situation when the systems are metallic.
The red and blue sites correspond to ${\bf S}_{1/2} =\pm {\bf S}$ magnetic sublattices. The green atom is non-magnetic, and its role is to alter the first-nearest neighbor Heisenberg (HEI) and Dzyaloshinskii-Moriya (DMI) interactions in its vicinity through the spin-orbit interaction. The green atom is in the plane with the red and blue sites. We also assume that there is no second-nearest neighbor HEI or DMI through the green atom. This is a simplification; however, our results below remain unchanged even if these interactions are non-zero, provided their magnitude differs from that of the corresponding interactions on bonds without the green atom.
Exchange Hamiltonian of an antiferromagnet model shown in Fig. (\ref{fig:fig1}) is
\begin{align}
H &= 
J\sum_{\langle ij \rangle} S_{i}^{z}S_{j}^{z} 
+ J \sum_{\langle ij \rangle} \cos(\theta_{ij}) \left( S_{i}^{x}S_{j}^{x} + S_{i}^{y}S_{j}^{y}  \right)
\nonumber
\\
&
+J\sum_{\langle ij \rangle} \nu_{ij} \sin(\theta_{ij}) \left( S_{i}^{x}S_{j}^{y} - S_{i}^{y}S_{j}^{x}  \right)
\nonumber
\\
&
+ \sum_{\langle\langle ij \rangle\rangle;\alpha}J^{\prime}_{ij;\alpha} S^{\alpha}_{i} S^{\alpha}_{j} 
+ \sum_{\langle\langle ij \rangle\rangle}D^{\prime}_{ij} [{\bf S}_{i}\times {\bf S}_{j}]_{z}, 
\label{modelExc}
\end{align}
where $J>0$ and $D\equiv J\sin(\theta)$ is the HEI and DMI between the first neighbors labeled by the $\langle ..\rangle$ in the sum. 
In Fig. (\ref{fig:fig1}) $\nu_{ij} = \pm1$ for the bond along the arrow. If the arrow on the $\langle ij \rangle$ bond is reversed, then $\nu_{ij} = -\nu_{ji}$. If there is no arrow on the bond then $\theta_{ij} = 0$. The signs are chosen in correspondence with the underlying spin-orbit coupling.
In addition we have added $J^{\prime}$ and $D^{\prime}$, which is the HEI and DMI correspondingly between the second-nearest neighbors, labeled by the $\langle\langle .. \rangle\rangle$.
We assume that $J^{\prime} <0$, i.e. it is of the ferromagnetic sign. Such a sign supports the N\'{e}el order. In addition to this, second-nearest neighbor interactions may be bond and sublattice dependent. The details of these interactions will be specified below for two different models shown in Fig. (\ref{fig:fig1}). We stress that there is no external magnetic field in the system.

In \cite{Moriya1960a,Moriya1960b,Shekhtamn1992,ZyuzinFietePRB2012} it was shown that spin-orbit coupling not only results in the DMI but also alters the corresponding HEI constant. For example, within the Hubbard model the HEI between the spins on a particular bond is $J = 4t_{1}^2/U$, where $t_{1}$ is the electron hopping amplitude (index $1$ labels the first-nearest neighbor hopping) and $U$ is the Hubbard on-site repulsion. If in addition there is a spin-orbit coupling on the bond, for example $\propto i\lambda_{1} \sigma_{z}$ (see \cite{ZyuzinFietePRB2012} for more details), where $\lambda_{1}$ is the amplitude of the spin-orbit coupling and $\sigma_{z}$ is defined by the geometry, i.e. when both electron hopping and a gradient of electric potential are in the same plane (in our case it is $x-y$ plane), then the HEI constant becomes $J= 4(t_{1}^2+\lambda_{1}^2)/U$ and the DMI between spins appears on this bond. 
As a result we may write the exchange Hamiltonain for such a bond as in \cite{ZyuzinFietePRB2012}, i.e.
\begin{align}
H_{ij} =
 JS_{i}^{z}S_{j}^{z} 
 &
+ J\cos(\theta)  \left( S_{i}^{x}S_{j}^{x} + S_{i}^{y}S_{j}^{y}  \right)\nonumber
\\
&
+J\sin(\theta) \left( S_{i}^{x}S_{j}^{y} - S_{i}^{y}S_{j}^{x}  \right),\label{DMI}
\end{align}
where $\theta = 2\arctan\left( \lambda_{1}/t_{1} \right)$ and the DMI constant is $D = J\sin(\theta)$. 
Hence the chosen model Hamiltonian given in Eq. (\ref{modelExc}).
Because of the out-of-plane DMI, the HEI in Eq. (\ref{modelExc}) is minimized by the N\'{e}el order in $z-$direction. 
We treat $J^{\prime}_{\alpha;ij}$ and $D^{\prime}_{ij}$ in Eq. (\ref{modelExc}) in the same fashion.
Where there is a DMI present, we write $J^{\prime}_{\alpha;ij} = J^{\prime}\cos(\zeta)$ for $\alpha = x,y$, $J^{\prime}_{z;ij} = J^{\prime}$ and $D^{\prime}_{ij} = J^{\prime}\sin(\zeta)$. Here $\zeta = 2\arctan\left( \lambda_{2}/t_{2} \right)$, where $t_{2}$ and $\lambda_{2}$ are the second-neareset hopping and spin-orbit coupling amplitude correspondingly.

We now assume that the N\'{e}el order is in $z-$direction and study magnons, which are fluctuations about the order.
Sublattices with ${\bf S}_{1}=+ {\bf S}$ and ${\bf S}_{2}=-{\bf S}$ spin will be called as the A/B sublattices correspondingly.
For our purposes it is convenient to use Holstein-Primakoff transformation from spins to boson operators. There are two sublattices and therefore there are two species of bosons describing the magnons. For A sublattice $S_{\mathrm{A}}^{+} = S_{\mathrm{A}}^{x}+iS_{\mathrm{A}}^{y}=\sqrt{2S-a^{\dag}a}a$, $S_{\mathrm{A}}^{-}=a^{\dag}\sqrt{2S-a^{\dag}a}$  and $S_{\mathrm{A}}^{z} = S-a^{\dag}a$, where $a^{\dag}$ and $a$ are boson operators. For B sublattice the transformation is slightly different, for example see \cite{ZyuzinKovalevPRL2016}, 
$S_{\mathrm{B}}^{+} =- S_{\mathrm{B}}^{x}+iS_{\mathrm{B}}^{y}=\sqrt{2S-b^{\dag}b}b^{\dag}$, $S_{\mathrm{B}}^{-}=b\sqrt{2S-b^{\dag}b}$  and $S_{\mathrm{B}}^{z} = -S+b^{\dag}b$, where $b^{\dag}$ and $b$ are boson operators. 

In the basis $\hat{\Psi}^{\dag}_{\bf k} = \left(a_{\bf k}^{\dag}, ~ b_{\bf k}^{\dag}, ~ a_{-\bf k}, ~b_{-\bf k}\right) $ the Hamiltonian of magnons is
\begin{align}\label{magnons}
\hat{H}_{\bf k} = 
SJ \left[
\begin{array}{cccc} 
3 + t_{{\bf k}}^{{\mathrm{A}}} & 0 & 0 & - \gamma_{\bf k}   \\
0 & 3 + t_{{\bf k}}^{{\mathrm{B}}}  & - \gamma_{-\bf k}  & 0 \\
0 & - \gamma_{-\bf k}^{*} & 3 + t_{-{\bf k}}^{{\mathrm{A}}}  & 0 \\
- \gamma_{\bf k}^{*} & 0 & 0 & 3 + t_{-{\bf k}}^{{\mathrm{B}}} 
\end{array}
\right].
\end{align}
Matrix element between the neighboring sites for the model I reads as
\begin{align}
\gamma_{\bf k} =  2 e^{i\frac{k_{x}}{2\sqrt{3}}}\cos\left( \frac{k_{y}}{2} \right) +  e^{-i\frac{k_{x}}{\sqrt{3}} +i\theta} ,
\end{align}
while for the model II it is
\begin{align}\label{gammaFerri}
\gamma_{\bf k} =  2e^{i\frac{k_{x}}{2\sqrt{3}}}\cos\left( \frac{k_{y}}{2} - \theta \right) + e^{-i\frac{k_{x}}{\sqrt{3}}} .
\end{align}
In obtaining these matrix elements of magnons within the sublattice, i.e. $t_{{\bf k}}^{{\mathrm{A}/\mathrm{B}}} $, 
we point out that due to the N\'{e}el order an extra minus sign in the amplitude of the second-nearest DMI appears within one sublattice as compared to another. 
For the model I, the matrix element within the sublattice A is
\begin{align}\label{tkAGenuine}
t_{{\bf k}}^{\mathrm{A}} &= 2t -  t\cos\left( \frac{\sqrt{3}k_{x}}{2} + \frac{k_{y}}{2} + \zeta \right)
-t\cos(k_{y}- \zeta) 
\end{align}
and within the sublattice B it is,
\begin{align}\label{tkBGenuine}
t_{-{\bf k}}^{\mathrm{B}} &= 2t -  t\cos\left( \frac{\sqrt{3}k_{x}}{2} - \frac{k_{y}}{2} + \zeta \right)
-t\cos(k_{y}+\zeta), 
\end{align}
where here and below $t=J^\prime/J$ which not to be confused with $t_{1/2}$ used above.
For the model II, the matrix elements are
\begin{align}
t_{{\bf k}}^{\mathrm{A}} &= 2t - 2 t \cos\left( \frac{\sqrt{3}k_{x}}{2} \right)\cos\left( \frac{k_{y}}{2} +\zeta \right) 
\end{align}
and
\begin{align}
t_{-{\bf k}}^{\mathrm{B}} &= t - t\cos\left( k_{y}+\zeta \right) .
\end{align}
We note that the second nearest neighbor Heisenberg exchange interaction is different as well as anisotropic in the two magnetic sublattices \cite{Naka2019,Pires} due to the position of the green atom. 

\begin{figure}[h] 
\includegraphics[width=0.4 \columnwidth ]{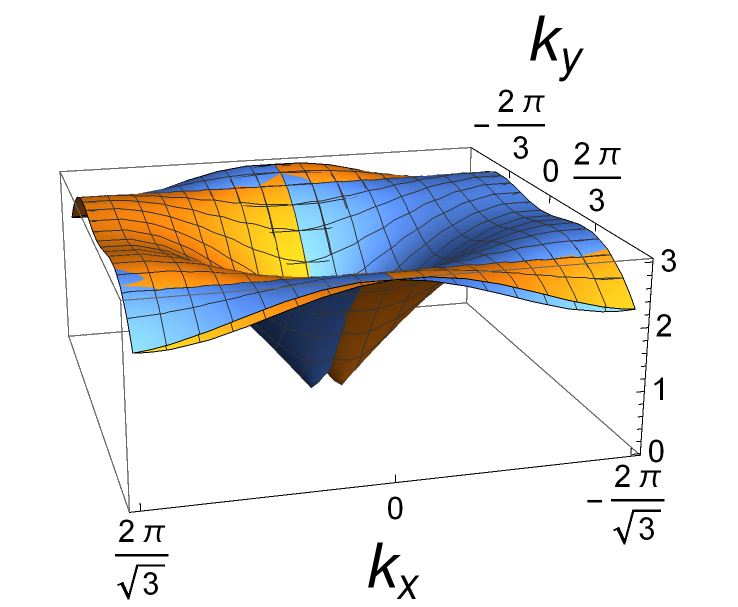} ~~
\includegraphics[width=0.4 \columnwidth ]{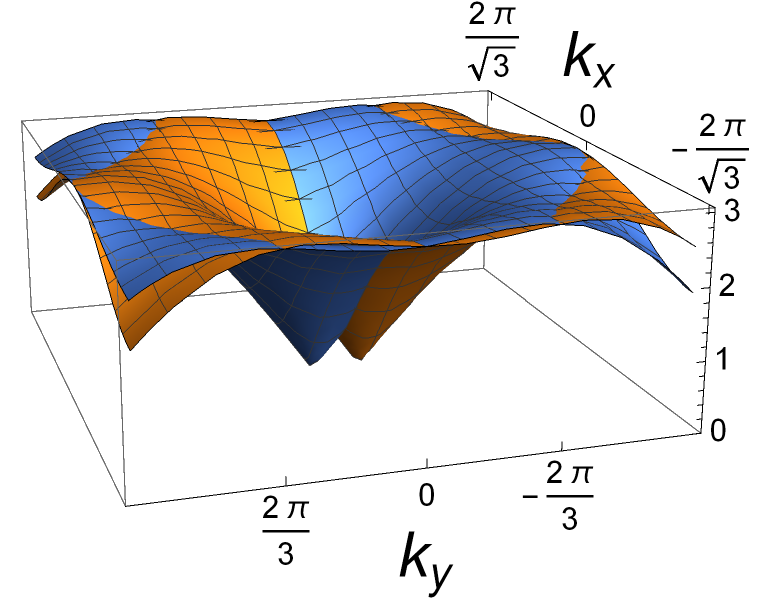} ~~

\protect\caption{Spectrum defined in Eq. (\ref{spectrum}) of the two antiferromagnetic magnon modes, plotted for $\theta = -0.2$ and $t=\zeta=0$ values. Left/right plots correspond to the model I/II. The plots are shown to emphasize the role of DMI in splitting of the magnon modes in momentum at the ${\bf \Gamma}$ point.}
\label{fig:fig2}  
\end{figure}

\begin{figure}[h] 
\includegraphics[width=0.4 \columnwidth ]{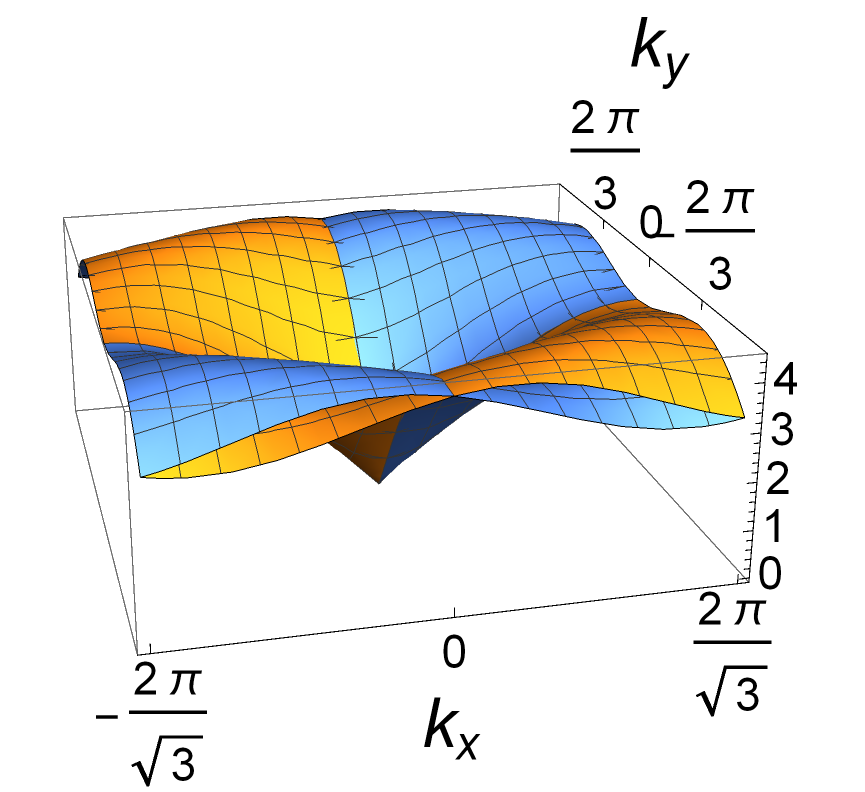} ~~
\includegraphics[width=0.4 \columnwidth ]{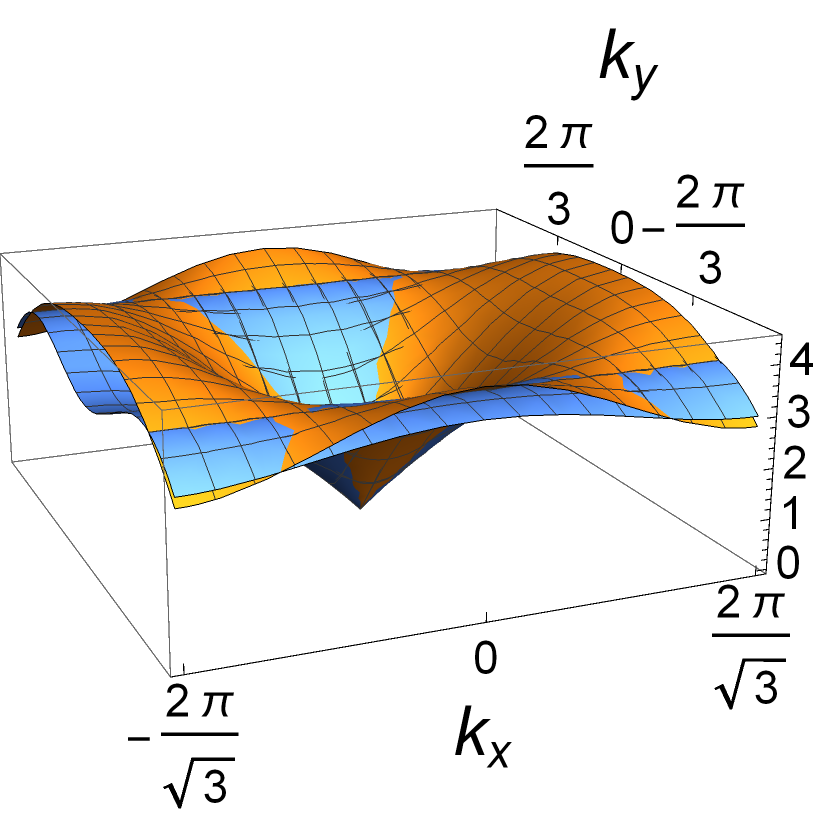} ~~

\protect\caption{Spectrum defined in Eq. (\ref{spectrum}) of the two antiferromagnetic magnon modes, plotted for $\theta = \zeta= 0.$ and $t=0.4$ values. Left/right plots correspond to the model I/II. The plots are shown to emphasize the role of the anisotropic exchange interaction given by $t$ in splitting of the magnon modes.}
\label{fig:fig3}  
\end{figure}

The solution to the eigenvalue equation $\hat{H}_{\bf k}\hat{\Psi} = \hat{\sigma}_{3} \hat{E}_{\bf k}\hat{\Psi}$ gives the magnon energy spectrum which is
\begin{align}
\hat{E}_{\bf k} = \mathrm{diag}\left[\epsilon^{(1)}_{{\bf k};+},~ \epsilon^{(2)}_{{\bf k};+},~\epsilon^{(2)}_{{\bf k};-},~\epsilon^{(1)}_{{\bf k};-}\right],
\end{align}
where $\hat{\sigma}_{3} = \mathrm{diag}[1,1,-1,-1]$ is the Pauli matrix in the magnon basis (see also \cite{comment1}).
We have defined
\begin{align}\label{spectrum}
\epsilon^{(1)}_{{\bf k};\pm} = 
SJ\left(\delta_{\bf k} \pm \varepsilon_{\bf k} \right),
~~~
\epsilon^{(2)}_{{\bf k};\pm} = 
SJ\left( -\delta_{-\bf k}\pm \varepsilon_{-\bf k} \right),
\end{align}
corresponding to two branches of antiferromagnetic magnons \cite{Auerbach}, where $\delta_{\bf k} = \frac{t_{{\bf k}}^{\mathrm{A}} - t_{-{\bf k}}^{\mathrm{B}}}{2}$,  $\varepsilon_{\bf k} \equiv \sqrt{(3+T_{\bf k})^2 - \vert \gamma_{\bf k}\vert^2}$ and $T_{\bf k} \equiv \frac{t_{{\bf k}}^{\mathrm{A}} + t_{-{\bf k}}^{\mathrm{B}}}{2}$ was introduced for convenience. 
In both models the first-nearest neighbor DMI entered the dispersion as a shift of the momentum, just like the vector potential in fermions does. The role of the DMI as the vector potential for ferromagnetic magnons has been studied in \cite{KovalevZyuzinLiPRB2017}.
Because of the DMI and presence of $T_{\bf k}$, the dispersion has the $\gamma_{\bf k}^{*}\neq \gamma_{-{\bf k}}$ and $\vert \gamma_{\bf k}\vert \neq \vert \gamma_{-\bf k}\vert$ properties in both models. We plot the dispersion of magnons in Fig. (\ref{fig:fig2}) for $t_{\bf k}^{\mathrm{A}/\mathrm{B}} = 0$. Indeed, the DMI breaks the degeneracy of the spin-up and spin-down magnon branches at the ${\bm \Gamma}$ point by shifting them in momentum. 
Broken degeneracy of the magnons modes \cite{Naka2019,Pires} by non-zero anisotropic second-nearest neighbor Heisenberg exchange interaction given by $T_{\bf k}$ in the spectrum Eq. (\ref{spectrum}) is shown in Fig. (\ref{fig:fig3}).
The spectrum obeys a $\epsilon_{{\bf k};\pm}^{(1)} = -\epsilon_{-{\bf k};\mp}^{(2)}$ property. 

\section{Magnon thermal Hall effect}
The thermal Hall effect of antiferromagnetic magnons has been derived in \cite{MatsumotoShindouMurakamiPRB2014} ( also see \cite{comment1,comment2}).
Magnon thermal Hall effect is defined as the heat current \cite{Luttinger1964} carried by the magnons, which flows in the transverse direction to the applied temperature gradient $\nabla_{\beta}\chi$, namely,
\begin{align}
j^{\mathrm{TH}}_{\alpha} 
= \frac{1}{2TV} \sum_{{\bf k};n} 
(\hat{\Omega}_{\alpha\beta;{\bf k}})_{nn}  c_{2}\left[ (\hat{E}_{\bf k})_{nn} \right]  \nabla_{\beta} \chi,
\end{align}
where $c_{2}(x) = \int_{0}^{x}dzz^2\frac{dg(z)}{dz}$, where $g(z) = (e^{z/T} - 1)^{-1}$ is the Bose-Einstein distribution function.
Berry curvature of magnons is a $4\times4$ diagonal matrix,
\begin{align}
\hat{\Omega}_{xy;{\bf k}} = \mathrm{diag}\left( \Omega_{xy;{\bf k}},~\Omega_{xy;-{\bf k}},~-\Omega_{xy;-{\bf k}},~-\Omega_{xy;{\bf k}} \right),
\end{align}
where
$
\Omega_{xy;{\bf k}} = \frac{w_{xy;{\bf k}}}{2\varepsilon_{\bf k}^{3}} 
$
and where
\begin{align}
w_{\alpha\beta;{\bf k}}&=  (3+T_{\bf k})
\left[ 
\partial_{\alpha}\mathrm{Im}\gamma_{\bf k} \partial_{\beta}\mathrm{Re}\gamma_{\bf k} 
- 
\partial_{\beta}\mathrm{Im}\gamma_{\bf k} \partial_{\alpha}\mathrm{Re}\gamma_{\bf k} 
\right]
\nonumber
\\
&
- \mathrm{Re}\gamma_{\bf k}
\left[ 
\partial_{\alpha}\mathrm{Im}\gamma_{\bf k} \partial_{\beta}T_{\bf k}
- 
\partial_{\alpha}T_{\bf k} \partial_{\beta}\mathrm{Im}\gamma_{\bf k} 
\right]
\nonumber
\\
&
+ \mathrm{Im}\gamma_{\bf k}
\left[ 
\partial_{\alpha}\mathrm{Re}\gamma_{\bf k} \partial_{\beta}T_{\bf k} 
- 
\partial_{\alpha}T_{\bf k} \partial_{\beta}\mathrm{Re}\gamma_{\bf k} 
\right].
\end{align}
It is instructive to introduce a part of the curvature
\begin{align}
\bar{w}_{\alpha\beta;{\bf k}}  = 
\partial_{\alpha}\mathrm{Im}\gamma_{\bf k} \partial_{\beta}\mathrm{Re}\gamma_{\bf k} 
- 
\partial_{\beta}\mathrm{Im}\gamma_{\bf k} \partial_{\alpha}\mathrm{Re}\gamma_{\bf k},
\end{align}
which is the part of the Berry curvature due to the first-nearest neighbor HEI and DMI.

We can arrange terms in the expression for the magnon thermal Hall effect as
\begin{align}
j^{\mathrm{TH}}_{\alpha} 
= \frac{1}{TV} \sum_{{\bf k};n} 
\Omega_{\alpha\beta;{\bf k}} \left[ c_{2}\left( \delta_{\bf k} + \varepsilon_{\bf k}\right) + c_{2}\left(- \delta_{\bf k} + \varepsilon_{\bf k}\right) \right]\nabla_{\beta} \chi.
\end{align}
From \cite{ZyuzinKovalevPRL2016} it is known that odd in momentum Berry curvature of antiferromagnetic magnons may be due to the lattice structure (honeycomb lattice), while from \cite{Pires} (checkerboard lattice with $\textit{d-}$wave DMI and finite $\delta_{\bf k}$) it is known that even in momentum Berry curvature may be due to the even in momentum DMI between the sublattices. However, both models, \cite{ZyuzinKovalevPRL2016} and \cite{Pires} don't allow for non-zero magnon thermal Hall effect because of the symmetry between the magnetic sublattices. These are genuine antiferromagnets and we are going to go over an example of such antiferromagnet in Section \ref{SectionGenuine}.
Then, in \cite{Mook,CommentAlter} an antiferromagent on the rutile lattice was shown to have finite THE due to the even in momentum Berry curvature, similarly to \cite{Pires}, originating from DMI. It appears that the antiferromagnet on a rutile lattice belongs to the symmetry class of Dzyaloshinskii's weak ferromagnets \cite{Dzyaloshinskii1958}, hence THE of magnons is expected. We will discuss THE of magnons in a collinear weak ferromagnet phase of our model in the Section \ref{SectionWeak}. 
In the Section \ref{SectionFerri}) we show that there is another possibility for non-zero magnon thermal Hall in collinear antiferromagnets shown in Fig. (\ref{fig:fig1}) with odd in momentum magnon Berry curvature due to nearest neighbor HEI, and with the even in momentum Berry curvature due to the nearest neighbor DMI. We will show that magnon THE will be non-zero only when all the symmetries between the two magnetic sublattices are broken, and hence the system is a ferrimagnet in this case. Let us discuss the details of models I and II.

\section{Mirror-symmetric genuine antiferromagnet}\label{SectionGenuine}
In the model I shown in the left of Fig. (\ref{fig:fig1}), the two magnetic sublattices are symmetric to each other.
The part of the Berry curvature due to nearest neighbor HEI is
\begin{align}
\bar{w}_{xy;{\bf k}} = \frac{1}{\sqrt{3}}\sin\left( \frac{k_{y}}{2} \right) 
\left[ \cos\left( \frac{k_{y}}{2} \right) - \cos\left( \frac{\sqrt{3}k_{x}}{2} +\theta \right) \right],
\end{align}
while 
\begin{align}
T_{\bf k} = 2t + t\cos\left( \frac{\sqrt{3}k_{x}}{2} +\zeta \right)\cos\left( \frac{k_{y}}{2}\right) + t \cos(\zeta) \cos\left( k_{y} \right).
\end{align}
It is clear that the integration over the momentum will make the THE to vanish in this model.
It is instructive to see how the Dzyaloshinskii-Turov symmetry \cite{Dzyaloshinskii1958,Turov1965} arguments support our result. We introduce a N\'{e}el vector ${\bf L} = {\bf S}_{1}-{\bf S}_{2}$ and the magnetization ${\bf M}$. The magnetization ${\bf M}$ may or may not be related to ${\bf S}_{1}+{\bf S}_{2}$. 
In the former case the magnetization is due to the canting of the N\'{e}el order, while in the latter case the magnetization is carried by magnons (or fermions if the system is metallic). For example, we can have ${\bf M} \neq 0$ even when ${\bf S}_{1}+{\bf S}_{2} = 0$. In this paper we are interested in exactly this scenario.
 We then analyze whether a Dzyaloshinskii's invariant of the $M_{\alpha}L_{\beta}$ (where $\alpha, \beta = x,y,z$) form can be realized in the system. For it to be non-zero, it has to be invariant under all symmetries of the system. If it is non-zero, it would mean that $L_{\alpha}$ can generate a finite magnetic moment $M_{\beta}$ in the system. In two dimensions we are expecting only $M_{z}$ to be generated. Both vectors transform as pseudovectors, and in addition, ${\bf L}$ changes sign if the magnetic sublattices are exchanged.
For $L_{y}$ and $L_{z}$, a reflection in the $y-z$ plane which crosses the vertical bond in the center is the symmetry connecting the two magnetic sublattices. Under this symmetry operation there is no Dzyaloshinskii's invariant of the $M_{z}L_{z}$ or $M_{z}L_{y}$ form. Indeed, $L_{z}$ and $L_{y}$ are invariant, while $M_{z}$ changes sign. It might appear that a symmetry of $\pi$ rotation about the center of the hexagon and a reflection in the $x-z$ plane which crosses the center of the hexagon and time-reversal allows for the $M_{z}L_{x}$ from the list of Dzyaloshinskii's invariants
However, a reflection in the $x-y$ plane and time-reversal operation eliminates the $M_{z}L_{x/y}$ combinations. All in all, the system in the left of Fig. (\ref{fig:fig1}) is a genuine antiferromagnet with zero magnetic moment, and magnon THE is not expected in this system.

\section{Collinear ferrimagnet}\label{SectionFerri}
\begin{figure}[h] 
\includegraphics[width=0.58 \columnwidth ]{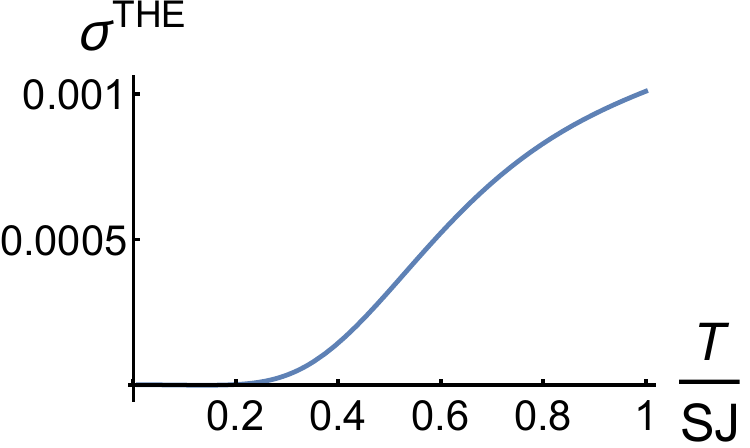} ~~~~~
\includegraphics[width=0.25 \columnwidth ]{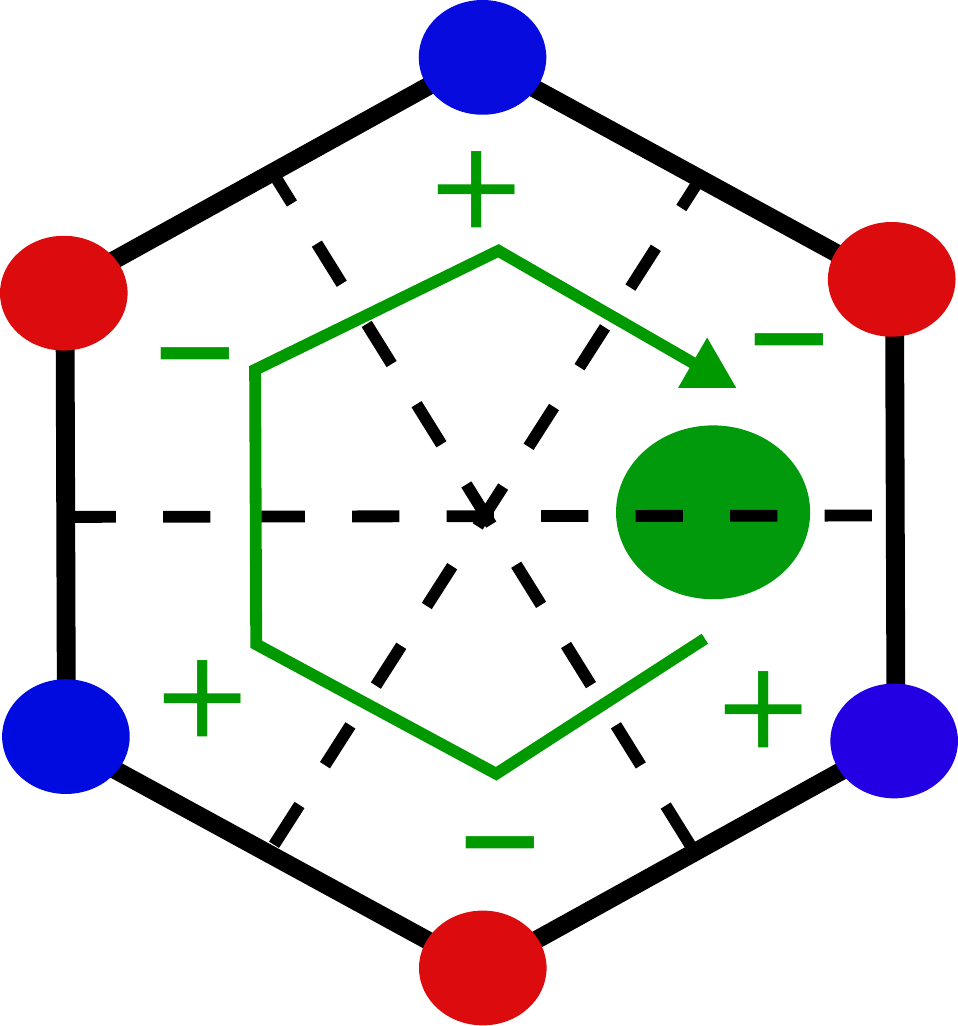}

\protect\caption{The N\'{e}el order is in $z-$direction. Left: THE of magnons for the model B for $\theta = 0.2$, $t=0.4$, $\zeta =0$. We checked that $\theta \rightarrow -\theta$ will change the sign of the THE. Right: evolution of the sign of the THE with the position of the green atom. }
\label{fig:fig4}  
\end{figure}

We now turn our attention to the model II shown in the right of Fig. (\ref{fig:fig1}). Now, there are no symmetries between the two magnetic sublattices in this model.
The part of the Berry curvature due to nearest neighbor HEI is
\begin{align}\label{BerryB}
\bar{w}_{xy;{\bf k}} = \frac{1}{\sqrt{3}}\sin\left( \frac{k_{y}}{2} -\theta \right) 
\left[ \cos\left( \frac{k_{y}}{2} -\theta \right) - \cos\left( \frac{\sqrt{3}k_{x}}{2} \right) \right],
\end{align}
while 
\begin{align}\label{TB}
T_{\bf k} = \frac{3t}{2} + t\cos\left( \frac{\sqrt{3}k_{x}}{2} \right)\cos\left( \frac{k_{y}}{2} + \zeta \right) + \frac{t}{2}\cos\left( k_{y} + \zeta \right) ,
\end{align}
therefore, a combination $\bar{w}_{xy;{\bf k}}T_{\bf k}$ of the Berry curvature has a part $\propto t\sin(\zeta + 2\theta) \cos^2\left( \frac{\sqrt{3}k_{x}}{2} \right)\sin^2\left( \frac{k_{y}}{2}\right)$ which doesn't vanish upon angle integration.
We can conclude that THE will be non-zero if there is no symmetry between the two magnetic sublattices. Certain anisotropic HEI and DMI appear in the Hamiltonian of magnons as a result of broken symmetry between the sublattices. For example, in our model, amplitude of THE is defined by such anisotropic HEI and DMI within the sublattices or anisotropic HEI within the sublattices and DMI between the sublattices. We checked that there is no THE when $t=\zeta=0$ and $\theta \neq 0$, meaning that DMI between the sublattices can be gauged away from the dispersion of magnons in this case. Namely, although such a DMI seemingly may make the Berry curvature Eq. (\ref{BerryB}) not to vanish upon angle integration, a simple shift of the momentum $ \frac{k_{y}}{2} -\theta \rightarrow \frac{k_{y}}{2}$ removes $\theta$ from the Berry curvature making THE to vanish. However, such a shift doesn't fully gauge away the DMI $\theta$ when $\cos\left( \frac{k_{y}}{2} + \zeta \right)$ is present in Eq. (\ref{TB}) thus making THE to be non-zero.  We plot $\sigma^{\mathrm{THE}}$ defined as  $j^{\mathrm{TH}}_{\alpha} 
= \sigma^{\mathrm{THE}}\nabla_{\beta} \chi({\bf r})$ in the left of Fig. (\ref{fig:fig4}) for $\theta = 0.2$, $t = 0.4$ and $\chi=0$ as a function of temperature. We have checked that finite $\chi$ also results in the THE of magnons.

A symmetry argument \cite{Dzyaloshinskii1958,Turov1965} can also be applied to the model shown in the right of Fig. (\ref{fig:fig1}). 
We consider the case when the green atom is in the same plane with the red and blue atoms.
Finite $M_{z}$ can't be generated for the $L_{x}$ and $L_{y}$ N\'{e}el orders. 
It does appear that a reflection in the $x-z$ plane and time-reversal allows for the $M_{z}L_{x}$ invariant, however, a reflection in the $x-y$ plane and time-reversal operation changes sign of the $M_{z}L_{x/y}$ combinations. In addition, $M_{z}L_{y}$ is eliminated by reflection in the $x-z$ plane. A $M_{z}L_{z}$ combination, on the other hand, is invariant under the reflection in the $x-y$ plane and time-reversal operation. The remaining symmetries, which are (i) reflection about the corresponding $y-z$ plane, $\pi$ rotation about the center of the hexagon, and time-reversal; (ii) reflection in the corresponding $x-z$ plane and time-reversal, don't change the sign of $M_{z}L_{z}$. In fact, these two symmetries are identical to each other. 
All in all, $M_{z}L_{z}$ is the only allowed Dzyaloshinskii's invariant, and thus magnon THE is expected in the system for the N\'{e}el order in $z-$ direction. Finally, we can make a claim that when all atoms are in the same plane, a reflection in the $x-y$ plane and time-reversal leaves $M_{z}L_{z}$ to be the only invariant in case when there is no symmetries between the magnetic sublattices. Such ferrimagnets can be called as the fluctuational-type (I-type) in which case it is only the fluctuations that result in the magnetization $M_{z}$. Indeed, no canting of the N\'{e}el order that is in $z-$direction, can result in finite magnetization in $z-$direction.

However, lifting of the green atom from the plane of the lattice would allow for the $M_{z}L_{x}$ invariant. This would distinguish such collinear ferrimagnets from the I-type ones. This is because now, in principle, the N\'{e}el order in $x-$direction can cant in the $z-$direction to produce finite magnetization. We note that, in addition, fluctuations can also result in finite magnetization in $z-$direction with the N\'{e}el order being intact. Therefore, such ferrimagnets can be called as the II-type. 
Analysis of the magnon THE in such ferrimagnets resemble that in the weak ferromagnets that will be discussed in Section \ref{SectionWeak}.

Let us draw a possible implementation of predicted in this paper scenario of non-zero magnon THE. Assume that charges of the green and red/blue atoms are of opposite signs. In our model, the most symmetric phase is when the green ion is in the center of the hexagon. No magnon THE is expected in this case. 
The system in the most symmetric phase is paraelectric, meaning there is no dipole moment in the unit cell. Then, by applying external electric field in the plane of the system, we can displace the green ion to take positions anywhere in the hexagon plaquette. For example in the way it is shown in Fig. (\ref{fig:fig1}).
As discussed in the vicinity of Eq. (\ref{DMI}), the green ion alters HEI and DMI due to the spin-orbit coupling it creates. Thus, the electric field can control values of HEI and DMI.
 Directions of the electric field corresponding to positions of the green ion as in the model I, left Fig. (\ref{fig:fig1}), will correspond to zero THE.  In addition to this position of the green ion, there are five more positions which don't break the symmetry between the magnetic sublattices.
Such positions are along a normal drawn from centers of the links of the hexagon.
Position of the green ion as in the Model II, right Fig. (\ref{fig:fig1}), and its mirrored position in the opposite corner of the hexagon, will have opposite in sign finite magnon THE. Thus, during the $2\pi$ rotation of the electric field in the plane of the system, the magnon THE will cross zero six times as shown in the right of Fig. (\ref{fig:fig4}).

\section{Collinear weak ferromagnet}\label{SectionWeak}
Having understood genuine antiferromagnet and ferrimagnetic systems, we now turn our attention to the weak ferromagnet\cite{BorovikRomanov1957,Dzyaloshinskii1958} case.
Weak ferromagnet is a N\'{e}el ordered antiferromagnet in which the symmetry allows for a finite magnetic moment.
In weak ferromagnets, contrary to ferrimagnets, the magnetic sublattices are connected to each other by some symmetry operation \cite{Dzyaloshinskii1958}.
Finite magnetic moment in weak ferromagnets might be due to the canting of the N\'{e}el order or be carried by magnons (or fermions in case the system is metallic). We consider the case when the N\'{e}el order is collinear, i.e. there is no canting of the order, and the magnetic moment is carried by the magnons. We stress that canting of the N\'{e}el order isn't necessarily for an antiferromagnet to be a weak ferromagnet. This is because it is rather the symmetry that distinguishes weak ferromagnets from other collinear antiferromagnets \cite{Dzyaloshinskii1958,Turov1965}.  
\begin{figure}[h] 
\includegraphics[width=0.8 \columnwidth ]{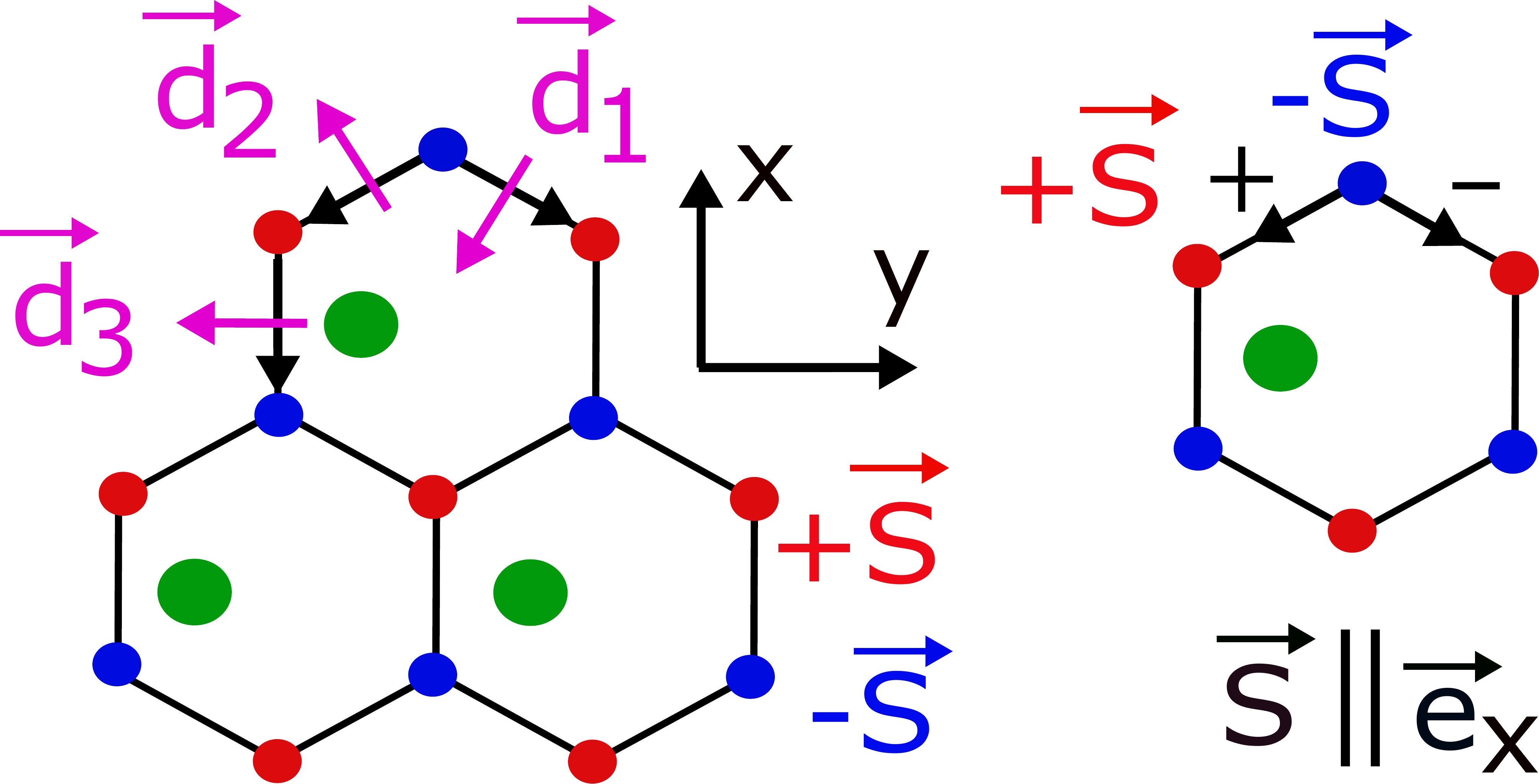} 

\protect\caption{The green atom is lifted from the $x-y$ plane making the system a weak ferromagnet with the $M_{z}L_{x}$ Dzyaloshinskii's invariant. Purple arrows are the directions of the DMI the green atom creates only due to the lifting from the $x-y$ plane, ${\bf d}_{1} = \frac{1}{2}(-\sqrt{3},-1)$, ${\bf d}_{2} = \frac{1}{2}(\sqrt{3},-1)$, ${\bf d}_{3} = (0,-1)$. In the right figure only the $x-$ component of the DMI, that is relevant for the N\'{e}el order in $x-$ direction, is shown. }
\label{fig:fig5}  
\end{figure}
\begin{figure}[h] 
\includegraphics[width=0.51 \columnwidth ]{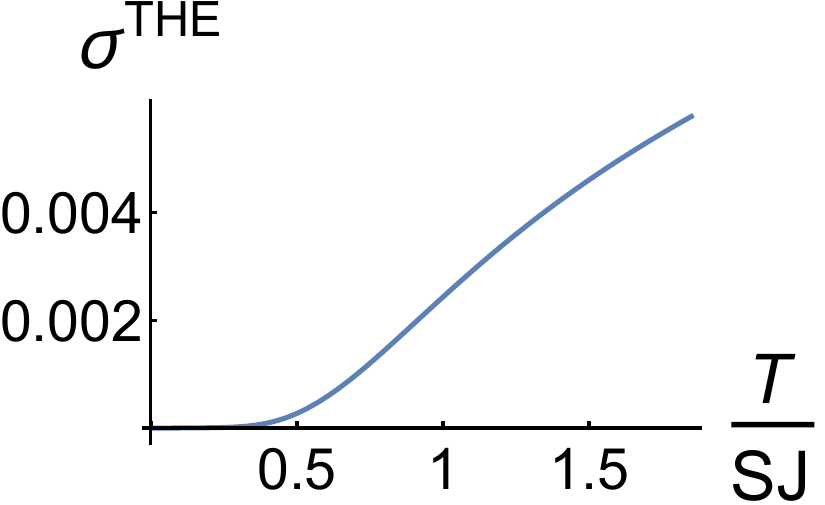} 

\protect\caption{THE of magnons for the weak ferromagnet described by Eq. (\ref{weak}) plotted for $\theta = 0.2$, $t=0.4$, $\zeta =0$. }
\label{fig:fig6}  
\end{figure}

A weak ferromagnet\cite{BorovikRomanov1957,Dzyaloshinskii1958} can be obtained from the genuine antiferromagnet, shown in the left of Fig. (\ref{fig:fig1}), by lifting the green atom from the $x-y$ plane. Lifting of the green atom allows for the $M_{z}L_{x}$ Dzyaloshinskii's invariant in the system.
We then set the N\'{e}el order in $x-$ direction and take into account only such DMI that are activated by a comination of the order in $x-$ direction and lifted green atoms. Such DMI is shown in Fig. (\ref{fig:fig5}) by ${\bf d}_{1} = \frac{1}{2}(-\sqrt{3},-1)$, ${\bf d}_{2} = \frac{1}{2}(\sqrt{3},-1)$, ${\bf d}_{3} = (0,-1)$. See Ref. \cite{KovalevZyuzinLiPRB2017} for an example of how DMI is defined with these vectors. When the N\'{e}el order is in $x-$ direction, only the $x$ component of ${\bf d}_{1/2/3}$ is relevant, and we can reduce the DMI only to these components. The plus and minus signs in the right figure in Fig. (\ref{fig:fig4}) is the reduced DMI.
The exchange Hamiltonian between the neighboring spins and the DMI due to the lifting of the green atom from the $x-y$ plane is 
\begin{align}
H_{\mathrm{weak}}=
\sum_{\langle ij \rangle}&\left[
  J S_{i}^{x}S_{j}^{x} 
+ J\cos(\theta_{ij})  \left( S_{i}^{y}S_{j}^{y} + S_{i}^{z}S_{j}^{z}  \right) \right]
\nonumber
\\
 &
 +\sum_{\langle ij \rangle} J\sin(\theta_{ij}) \left( S_{i}^{y}S_{j}^{z} - S_{i}^{z}S_{j}^{y}  \right).
 \label{weak}
\end{align}
We thus have magnons described by Eq. (\ref{magnons}), in which $\gamma_{\bf k}$ is defined in Eq. (\ref{gammaFerri}) and $t^{\mathrm{A}/\mathrm{B}}_{\bf k}$ are defined in Eqs. (\ref{tkAGenuine}) and (\ref{tkBGenuine}). Therefore, magnons describing weak ferromagnet are a mixture of those describing genuine antiferromagnet and ferrimagnet. We stress that in the description of magnons the DMI in the weak ferromagnet is identical to the ferrimagnetic case. THE of magnons for the weak ferromagnet Eq. (\ref{weak}) is plotted in Fig. (\ref{fig:fig6}). There are more weak ferromagnets in Nature than genuine antiferromagnets \cite{Turov1965}. This maybe understood from the examples used in this paper. Genuine antiferromagnet given by the model A shown in the left of Fig. (\ref{fig:fig1}) is unstable towards formation of the weak ferromagnet because the in-plane position of the green atom is rather a fine tuned situation. Indeed, even a slight lifting of the position of the green atom from the plane of the crystal will turn genuine antiferromagnet into a weak ferromagnet. 
Realistic weak ferromagnets are presented by RuO$_2$, CrSb, CoF$_2$, NiF$_2$, $\alpha$-Fe$_2$O$_3$, MnTe, LuFeO$_3$, and many more. If one wants to undertand whether a given antiferromagnet belongs to the symmetry class of a weak ferromagnet, one has to look for the symmetry group of the system at hand and consult the table 3 in \cite{Turov1965}.
Presence or absence of the magnetic moment in these materials depends on the orientation of the N\'{e}el order in accord with the Dzyaloshinskii's invariant. For example, in the studied here system shown in Fig. (\ref{fig:fig5}) only the $L_{x}$ can generate finite magnetic moment $M_{z}$. Clearly, THE of magnons will be absent if the magnetic moment isn't allowed for certain directions of the N\'{e}el order.

\section{Conclusions}

To conclude, we have studied theoretical models, depicted in Fig. (\ref{fig:fig1}) and Fig. (\ref{fig:fig5}), of insulating compensated (total spin on the two sublattices is zero) collinear antiferromagnets that show magnon thermal Hall effect. We showed that for the magnon thermal Hall effect to be non-zero, the symmetry between the sublattices of an antiferromagnet must be broken. The antiferromagnet in this case is a collinear ferrimagnet.
We then showed that finite magnon thermal Hall effect can occur also in the case when the two magnetic sublattices are connected by some symmetry operation. The antiferromagnet in this case is a collinear weak ferromagnet \cite{Dzyaloshinskii1958,Turov1965}. 
We have corroborated our results by symmetry argument consistent with Dzyaloshinskii's classification scheme of antiferromagnets \cite{Turov1965}. In particular we have deduced Dzyalshinskii's invariant for every studied systems and showed that calculated finite magnon thermal Hall effect in collinear ferrimagnet and collinear weak ferromagnet is consistent with the invariants.

Furthermore, we have obtained two types of ferrimagnets, both depicted in the right of Fig. (\ref{fig:fig1}). The first one (type-I) is with the $M_{z}L_{z}$ Dzyaloshinskii's invariant. This means that no canting of the N\'{e}el order $L_{z}$ can result in $M_{z}$, and finite $M_{z}$ in the system is purely due to the magnons. 
In this system the green atom shown in the right of Fig. (\ref{fig:fig1}) is in the plane with the red and blue atoms.
Second type (II-type) of ferrimagnets is with the $M_{z}L_{x}$ Dzyaloshinskii's invariant. Such a ferrimagnet can have canting of the N\'{e}el order that would produce finite $M_{z}$ in addition to the magnon mechanism. In this ferrimagnet the green atom depicted in the right of Fig. (\ref{fig:fig1} is lifted from the plane of the lattice. Weak ferromagnets, on the other hand, always have an invariant with components of ${\bf M}$ and ${\bf L}$ being orthogonal to each other. For example in the Fig. (\ref{fig:fig5}) a weak ferromagnet with $M_{z}L_{x}$ invariant is shown.

We found magnon spin-splitting at the ${\bm \Gamma}$ point shown in Fig. (\ref{fig:fig2}) due to the Dzyaloshinskii-Moriya to be one of the ingredients for non-zero magnon thermal Hall effect. Another ingredient is the magnon spin-splitting away from the ${\bm \Gamma}$ point that is due to the anisotropic second-nearest Heisenberg exchange interaction given by $T_{\bf k}$ in Eq. (\ref{spectrum}) and shown in Fig. (\ref{fig:fig3}). 
We suggested a scenario when external electric field applied in the plane of the system can change magnon thermal Hall effect by altering symmetry of the lattice.

The author thanks A.A. Kovalev for discussions.
The author is grateful to Pirinem School of Theoretical Physics for hospitality during the Summer of 2025. This work is supported by FFWR-2024-0016.


\begin{references}
\bibitem{CommentAlter} It turns out altermagnets belong to the symmetry class of weak ferromagnets.
\bibitem{comment2} In \cite{MatsumotoShindouMurakamiPRB2014} and \cite{ZyuzinKovalevPRL2016} notations of the gravitational potential are slightly different from each other but the meaning is the same. In \cite{MatsumotoShindouMurakamiPRB2014} the temperature in the system is defined as $T({\bf r}) = T[1-\chi({\bf r})]$ and, thus, the gravitational field is $\propto \frac{1}{2}\{\frac{{\bf r}\cdot{\bm \nabla}\chi}{T},\hat{H}\}$, where $\chi({\bf r})$ is expanded to the linear order in ${\bf r}$. In \cite{ZyuzinKovalevPRL2016} the gravitational field was defined as $\propto \frac{1}{2}\{\frac{{\bf r}\cdot {\bm \nabla} T({\bf r})}{T},\hat{H}\}$, i.e. $T({\bf r})$ instead of $\chi({\bf r})$. In addition in \cite{ZyuzinKovalevPRL2016} a notation $\chi({\bf r}) \equiv \frac{T(\bf r)}{T} $ was used.
\end{references}
\end{document}